# Deep learning denoising for EOG artifacts removal from EEG signals

[1]Najmeh Mashhadi*, [2]Abolfazl Zargari Khuzani, [3]Morteza Heidari, [4]Donya Khaledyan
[1]The Department of Computer Science and Engineering, University of California, Santa Cruz, USA
[2]The Department of Electrical and Computer Engineering, University of California, Santa Cruz, USA
[3]School of Electrical & Computer Engineering, University of Oklahoma, Norman, USA
[4]Faculty of Electrical Engineering, Shahid Beheshti University, Tehran, Iran
* nmashhad@ucsc.edu

*Abstract*— **There are many sources of interference encountered in the electroencephalogram (EEG) recordings, specifically ocular, muscular, and cardiac artifacts. Rejection of EEG artifacts is an essential process in EEG analysis since such artifacts cause many problems in EEG signals analysis. One of the most challenging issues in EEG denoising processes is removing the ocular artifacts where Electrooculographic (EOG), and EEG signals have an overlap in both frequency and time domains. In this paper, we build and train a deep learning model to deal with this challenge and remove the ocular artifacts effectively. In the proposed scheme, we convert each EEG signal to an image to be fed to a U-NET model, which is a deep learning model usually used in image segmentation tasks. We proposed three different schemes and made our U-NET based models learn to purify contaminated EEG signals similar to the process used in the image segmentation process. The results confirm that one of our schemes can achieve a reliable and promising accuracy to reduce the Mean square error between the target signal (Pure EEGs) and the predicted signal (Purified EEGs).**

Keywords: ***EEG signals, denoising signal, EOG artifact removal, U-net structure, convolutional neural network ( CNN ).***

## I. Introduction

Electroencephalography (EEG) is a non-invasive method to record the brain's electrical activities over a period of time with the electrodes placed on the scalp. EEG records get contaminated by some biological artifacts, and among them, eye movements, which are recorded by electrooculograms, obscure the EEG signals more than other artifacts. In fact, the amplitude of ocular artifacts is quite large compared to the EEG signals' amplitude. Therefore, purifying the EEG signals from EOG signals is very important to interpret EEGs correctly[1-3]

There are different approaches to remove Ocular artifacts from contaminated EEG signals. One method is independent component analysis (ICA) [1,3,4], which decomposes multiple EEG channels into an equal number of independent components (ICs). Then, ICs that contain ocular artifacts are removed, and pure EEG from the remaining ICs is reconstructed [1]. This method causes information loss and needs a large number of EEG channels to identify all distinct characteristics of ocular artifacts in different ICs [3]. Another method is Empirical mode decomposition (EMD) that decomposes the signal into waveforms modulated in both amplitude and frequency and needs a reference signal for decomposition.

EMD is applied to a univariate and real-valued signal, while the bivariate EMD (BEMD) is suitable for complex signals [5]. In the two-step BEMD method, fractional Gaussian noise is used as a reference to preprocess the EOG signal in the first step. Then, the preprocessed EOG signal is utilized in the second step as a reference to clean the EEG signal [5]. The other widely used technique is the wavelet transform based on thresholding function [2-4, 6] that decomposes the contaminated EEG signals into sub-bands, and the coefficient of ocular artifacts are corrected based on a thresholding function in each sub-band. Then, the modified EEG signals are reconstructed based on the corrected coefficients. The wavelet transform does need a reference signal that makes it applicable in online EOG removal [6]. However, determining a proper threshold function that does not filter out the useful parts of the EEG signal and, at the same time, removes a certain ocular artifact is difficult [7]. Some other techniques are proposed by combining the mentioned methods resulting in a better performance compared to the previous works. For example, the wavelet transform and ICA jointly reduce the loss of residual informative data [3]. A new scheme using a correlation-based system and discrete wavelet transform (DWT) rule, achieves more accurate results than ICA and also removes common-line artifacts mixed with EOG artifacts [8]. For EOG artifact rejection, a few approaches are using a neural network. One of them is the wavelet neural network (WNN) method in which the thresholding function, used in the wavelet transform, is replaced with a neural network. WNN removes EEG artifacts effectively without diminishing useful EEG information even for very noisy datasets [6]. The other developed scheme uses a deep learning network (DLN) with three hidden layers. In this method, first, the sample pure EEG signals are used to train the network. Then, the trained DLN is used as a filter to automatically remove ocular artifacts from the contaminated EEG signals [1]. The neural network methods do not need additional EOG records as a reference in real-time EOG rejection, and there is no limit in the number of channels.

In this paper, we propose three deep learning-based schemes to purity EEG signals from EOG artifacts. In the first and best scheme, the contaminated EEG signals, captured from each electrode, are reshaped and converted to a grayscale image. The grayscale images of all samples for each electrode are used to train a U-Net model. Since 19 electrodes record our EEG dataset from 54 samples, we trained 19 U-NETs separately. During the training process, the predicted outputs are compared with the pure EEG signals recorded in a closed-eye experiment aiming to yield accurate predictions. We use the mean square error (MSE) as the loss function in the training process and as a function to evaluate our method performance. Results indicate that the predicted signals follow the corresponding target signals with low error.

## II. MATERIALS AND METHODS

### A. Dataset

We use a semi-simulated EEG dataset in which artifact-free EEG signals are manually contaminated with ocular artifacts using a realistic head model [9]. The significant feature of this dataset is that it contains the pure EEG signals with their corresponding contaminated signals. This dataset consists of the EEG signals of 54 participants recorded during a closed-eye experiment; so, the obtained EEGs are free of ocular artifact. During this experiment, the EGG signals were recorded by 19 electrodes placed based on the International 10–20 system. These electrodes are labeled FP1, FP2, F3, F4, C3, C4, P3, P4, O1, O2, F7, F8, T3, T4, T5, T6, Fz, Cz, and Pz. Based on International 10–20 system, the label of each electrode indicates the area of the brain from where the signals are being read (pre-frontal (Fp), frontal (F), temporal (T), parietal (P), occipital (O), and central (C)). The location of the other three electrodes (Fz, Cz, and Pz) are on the midline sagittal plane of the skull called z site.

In naming the labels, even numbers show that the electrode is placed on the right side of the head, whereas odd numbers refer to electrodes placed on the left side of the head [9]. Another experiment was performed on the same participants but with the eye-opened condition, and six electrodes around their eyes recoded the EOG signals. The difference between the value of upper and lower electrodes and the value of right and left electrodes creates vertical EOG (VEOG) and horizontal EOG (HEOG) datasets, respectively. Both experiments were performed around 30 seconds, and the sampling frequency is 200 Hz. The frequency of pure EEG signals is in the rage of 0.5-40 Hz, and the frequency of EOG signals is in the rage of 0.5-5 Hz. The contaminated (semi-simulated) EEG signals were obtained by applying the VEOG and HEOG on the pure signals based on a realistic contamination model as follows [9]:

$$Contaminated\ EEG_{i,j} = Pure\ EEG_{i,j} + a_j VEOG + b_j HEOG \quad (1)$$

The pure and contaminated EEG datasets consist of 54 matrices, and each matrix corresponds to a participant. The matrices have 19 rows so that each row has the records of each channel (electrode). The number of columns is in the range 5600 to 8400, which is different for each participant. Since the data are sampled with constant frequency, each person's different amount of collected data refers to different recording times.

### B. U-net structure

Nowadays, neural networks and deep learning models are important parts of detection, prediction, classification, segmentation, and recognition systems with different applications [10-28]. U-Net is a convolutional neural network (CNN) architecture used for accurate and fast image segmentation [10]. In this study, we use the U-NET model as a function to purify the contaminated EEG signals. As seen in Figure 1, the origin of the U-Net name is inspired by its structure's symmetric shape. U-Net is built on an architecture that uses a fully convolutional network (FCN) to transform image pixels into pixel class labels. The U-Net structure consists of three parts: the contracting (down-sampling) path, bottleneck, and expanding (up-sampling path). The contacting path comprises of four blocks, and each block holds two 3x3 convolution layer followed by an activation function and a 2x2 max pooling. The number of feature maps doubles at each pooling, starting with 64 feature maps for the first block [10]. In this paper, we use the exponential linear unit (ELU) as an activation function. ELU has a better performance when the batch normalization can not be applied in the training process due to the small number of data. The second part, bottleneck, is placed between the contracting and expanding paths and built from two convolutional layers without dropout. The last part is the expanding path includes four blocks, and each block contains 1) a deconvolution layer with stride two, which halves the number of feature channels 2) concatenation with the corresponding cropped feature map from the contacting path, 3) two 3x3 convolution layers followed by an activation function. The up-sampling path enables accurate localization combined with contextual information from the down-sampling path. In the last layer, there is a 1x1 convolution to map each 64-component feature vector to a certain number of classes. Totally, the network has 23 convolutional layers [10].

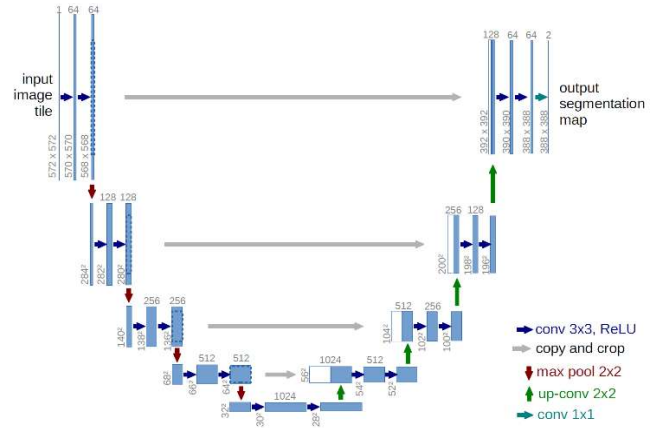

Fig. 1: U-net structure [10]

### C. Data preparation process

*a) Method 1:* The data preparation process used in this method is shown in Figure 2. Since the minimum length of all channels in our dataset is equal to 5400, we choose the first 5400 data points in each channel. A low-pass filter is then to discard a part of high-frequency noise from both pure and contaminated EEG signals. Applying this low-pass filter, positively impact the accuracy of our deep learning model. In the next step, we resample the data points to 6400 new samples aiming to convert them to an 8-bit grayscale image with a size of 80 x 80 pixels. During the grayscale conversion, the Min-Max Normalization is used to bring all data values in the range of 0 - 255.

The Min-Max Normalization is a simple normalization strategy which transforms $x$ to $y = f(x)$, so that if $(x_{max}, x_{min})$ are the min and max boundary of $x$ values, then $y_{min} = f(x_{min}), y_{max} = f(x_{max})$ is the min and max boundary of $y$. The best and most important advantage of Min-Max Normalization is that it preserves the relationship between the original data values, so we can use it without being worried about the correctness of our comparisons on transformed data. This transformation is as follows:

$$f(x) = \frac{x - x_{min}}{x_{max} - x_{min}} (y_{max} - y_{min}) + y_{min} \qquad (2)$$

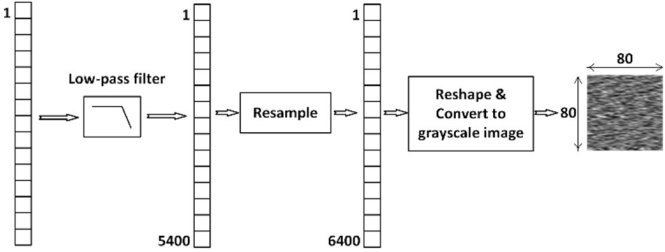

Fig. 2: data preparation process used in method 1

Since each channel of EEGs is captured from a different part of the brain and also affected by different kinds of artifacts, we train separate deep U-NET models in our first proposed scheme. As illustrated in Figure 3, we train 19 seperate U-NET models using the prepared grayscale image dataset of each channel.

*b) Method 2:* In the second proposed method, we collect the grayscale image datasets of all channels (generated by the data preparation technique in the first method) into a single image dataset. Therefore, the new dataset consists of 19 x54 grayscale images with a size of 80 x 80. This larger dataset is then fed to one single U-NET model for the training process, as shown in Figure 4.

*c) Method 3:* In the third method, after applying the low-pass filter to each channel, we put all the filters' outputs in a row to create a matrix of 5400 x 19, as shown in Figure 5. Then, we reshape and convert it to an 8-bit grayscale image with the new size of 512 x 256. In other words, all data channels are converted to a single grayscale image. The created dataset with 54 grayscale images of size 512 x 256 is used to train a single U-NET model, as seen in Figure 6.

*D. Training process*

We employ the same training algorithm and hyperparameters for all three proposed methods. Before running the training process, all image datasets are categorized into three sets, including training set, validation set, and test set with a ratio of 70%,15%, and 15%, respectively. Also, we set the training hyperparameters as follows in table 1.

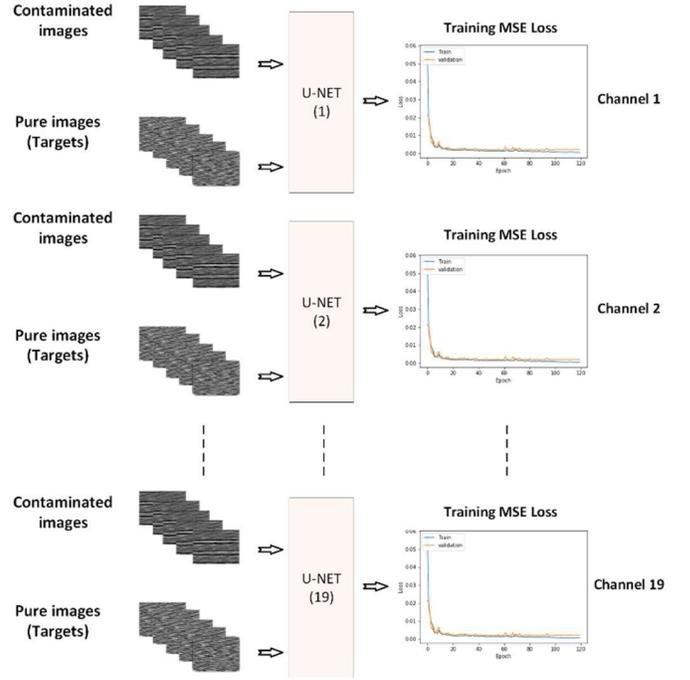

Fig. 3: U-net models used in method 1; 19 distinct U-net models are trained for 19 different channels.

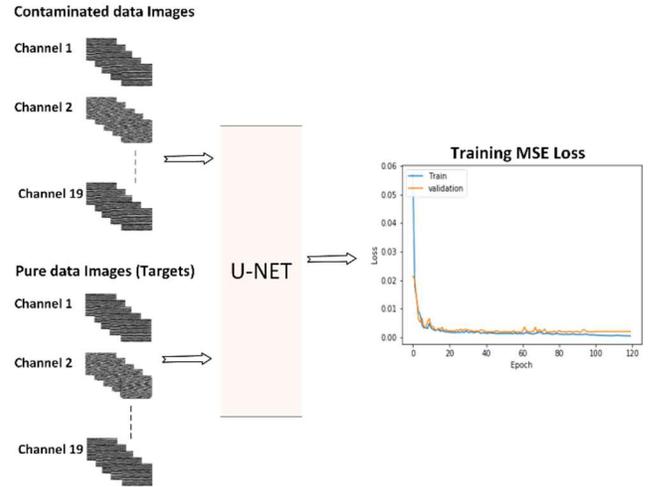

Fig. 4: U-net model used in method 2. The prepared grayscale images of all channels are passed through one U-NET model.

We use the Adaptive Moment Estimation (Adam) optimizer in the training process to minimize the loss function. Adam employs Momentum and Adaptive Learning Rates to converge faster and has a noticeably better performance than stochastic Gradient Descent to avoid getting stuck in local minimums and saddle points. Adam works well in practice and outperforms other Adaptive techniques. Besides, during the training algorithm, we apply two regularization techniques, drop out and early stopping, to improve our network performance.

Regularization techniques penalize certain model parameters if they're likely to cause overfitting. Dropout is a technique applied to neural networks that randomly sets some of the neurons' outputs to zero during training. This forces the network to better learn representations of the data by preventing complex interactions between the neurons: Each neuron needs to learn useful features. Early stopping will stop training when the validation score stops improving, even when the training score might be improving. This prevents overfitting on the training dataset. To evaluate the trained model performance after the training process, we use the test image dataset.

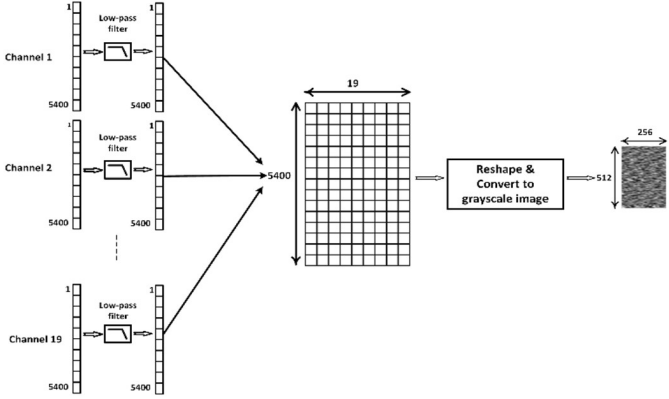

Fig. 5: U-net models used in method 3. All channel data points for each patient are concatenated into one grayscale image and passed through one U-NET model.

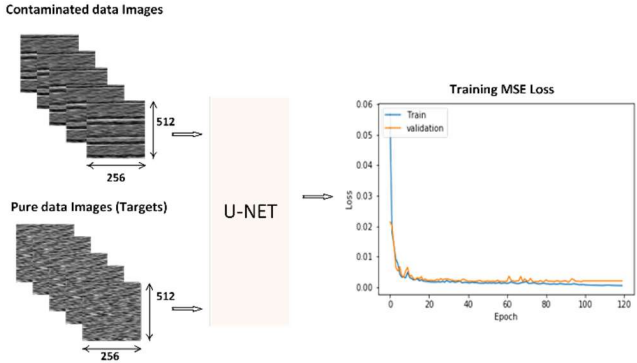

Fig. 6: U-net models used in method 3

Table1: training hyperparameters

| Parameter | Value |
|---|---|
| Learning rate | $10^{-4}$ |
| Training batch size | 2 |
| Max epoch | 120 |
| Training step per epoch | 19 |
| Optimization method | Adam optimizer |
| Dropout | 0.5 |
| Early stopping patient | 10 |
| Loss function | MSE |

## III. RESULTS

### A. Training result

The training and validation loss graphs of the three proposed learning schemes are shown in Figure 6. In the first learning scheme in which 19 distinct networks have been trained, the lowest training and validation loss is obtained from the 10th channel. As seen in figure 6, the lowest training and validation loss belongs to the first scheme (channel 10), while the second learning scheme results in the highest training and validation loss.

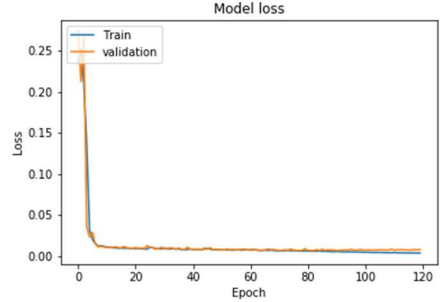

a) First learning scheme (channel 10)

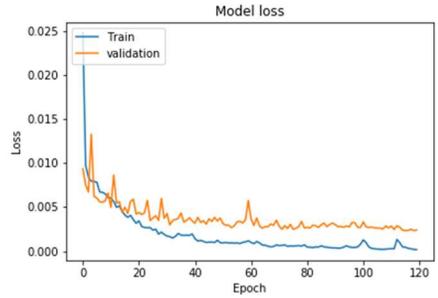

b) Second learning scheme

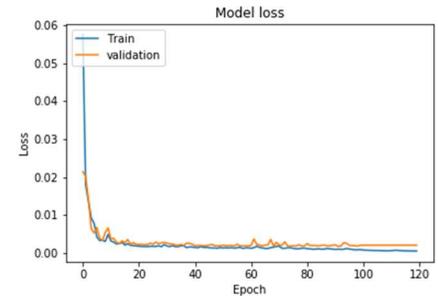

c) Third learning scheme

Fig. 6: Model loss graphs of the three proposed learning schemes

### B. Test result

We pass the test set images through the three trained networks to evaluate the trained models' accuracy in purifying the contaminated EEGs. Then, the Mean square Error (MSE) of the predicted outputs and their corresponding target signals (pure EEGs) are calculated. Table 2 presents the MSE for each channel obtained by applying the first proposed method. Table 3 presents the Average MSE values of the first and the MSE of the second and third models, respectively. As seen in Table 3, the first method achieves a lower MSE and higher accuracy than the other two. Besides, the trained model of channel 10 indicates the minimum MSE value among all other channels.

Table 2: MSE for each channel obtained by applying the first method and test set images

| Channel number | Mean Square Error |
|---|---|
| 1 | 0.00215 |
| 2 | 0.00128 |
| 3 | 0.00061 |
| 4 | 0.00137 |
| 5 | 0.00044 |
| 6 | 0.00049 |
| 7 | 0.00025 |
| 8 | 0.00016 |
| 9 | 0.00015 |
| 10 | 0.00010 |
| 11 | 0.00158 |
| 12 | 0.00103 |
| 13 | 0.00053 |
| 14 | 0.00072 |
| 15 | 0.00026 |
| 16 | 0.00016 |
| 17 | 0.00080 |
| 18 | 0.00039 |
| 19 | 0.00016 |

Table 3: The average MSE obtained by applying the first method and the MSEs achieved by applying the second and third methods

| Learning scheme | |
|---|---|
| Method 1 | Average MSE = 0.000573 |
| Method 2 | MSE = 0.0358 |
| Method 3 | MSE = 0.00712 |

Figure 7, as a sample, compares the target (pure) signal with predicted (purified) and contaminated signals obtained from the first proposed machine learning scheme for the channels 1, 4, and 11 in which the ocular artifacts are noticeable. As shown, our first method denoises the EEGs with high accuracy, and the predicted signals are following the pure target signals with an error close to zero.

## IV. CONCLUSION AND DISCUSSION

In this paper, we utilized a deep learning-based method to purify contaminated EEG signals from ocular artifacts. We had two datasets of pure EEG signals and contaminated EEGs obtained from 54 participants in a clinical experiment. We implemented the two-dimensional U-net structure, which is usually used in image segmentation, but we employed it for denoising purposes by converting the EEG signals to 2D grayscale images. Each channel in our dataset is data points recorded from a specific part of the brain according to the "International 10-20 system". In the next step, we developed three different methods to use the prepared grayscale images for training deep U-NET models. In the first method, we trained one U-net network separately for each channel of 19 channels and created image datasets (images with a size of 80 x 80). In the second method, just one single U-net network was trained for all 19 channel images (images with a size of 80 x 80). In the third method, first, we combined the dataset of all channels into a single dataset and then converted and reshaped samples into the larger grayscale images (images with a size of 512 x 256) to train one single U-net network. All three methods removed the ocular artifacts from the contaminated EEGs with high accuracy. However, we achieved a more precise result in the first method with the average mean square error of 0.00057 (comparing the purified EEGs with the pure EEGs). We can conclude that the amount and strength of ocular artifacts are different in each channel. Thus, training 19 separate deep learning models enables us to denoise each EEG channel data more efficiently.

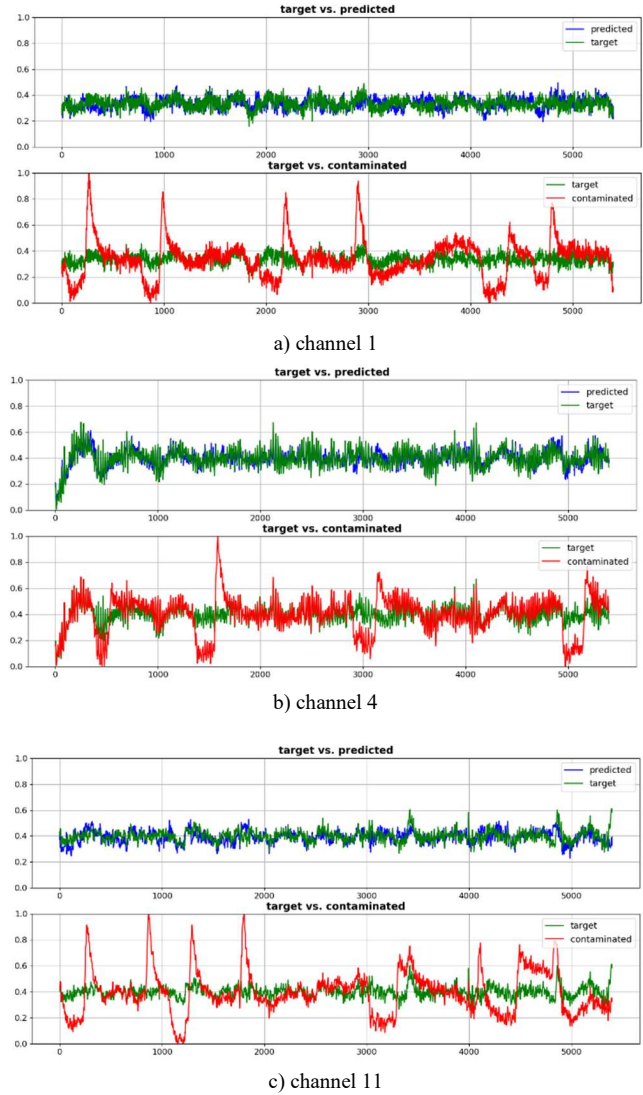

a) channel 1

b) channel 4

c) channel 11

Fig. 7: Target signal versus predicted signal (on the top) and Target signal versus contaminated signal (on the bottom) for the channels 1, 4, and 11